\documentclass[smallextended]{svjour3}
\usepackage{amsmath,amssymb,amsfonts}

\smartqed

\def\vol{\mathrm{vol}}

\title{%
Algebraic approach to quantum field theory on a class of noncommutative curved spacetimes}

\author{Thorsten Ohl \and Alexander Schenkel}
\institute{T.~Ohl \and A.~Schenkel \at Institut f\"ur Theoretische Physik und Astrophysik\\
  Universit\"at W\"urzburg\\
   Am Hubland \\
   D-97074 W\"urzburg, Germany\\
\email{ohl@physik.uni-wuerzburg.de}\\
 \email{aschenkel@physik.uni-wuerzburg.de}}
\date{December 2009}

\begin{document}
\maketitle


\begin{abstract}
In this article we study the quantization of a free real scalar field on a 
class of noncommutative manifolds, obtained via formal deformation quantization using 
triangular Drinfel'd twists. 
We construct deformed quadratic action functionals and compute the corresponding 
equation of motion operators. The Green's operators and the fundamental solution of the 
deformed equation of motion are obtained in terms of formal power series. It is shown that, using 
the deformed fundamental solution, we can define deformed $\ast$-algebras of field observables, which in 
general depend on the spacetime deformation parameter. 
This dependence is absent in the special case of Killing deformations, which include in particular the 
Moyal-Weyl deformation of the Minkowski spacetime.
\end{abstract}


\section{Introduction}
Quantum field theory (QFT) on noncommutative (NC) spacetimes is a subject of particular interest
 in modern mathematical and theoretical physics, see e.\,g.~\cite{Douglas:2001ba,Szabo:2001kg} for
 reviews. Even though some work on interacting QFTs on canonically deformed 
Euclidean or Minkowski space has been done using modified Feynman rules or other perturbative methods, the structure of free 
NC QFTs has still not been elucidated completely. For a collection of different approaches to QFT 
on Moyal-Weyl- or $\kappa$-deformed Minkowski spacetime see \cite{NCQFT} and references therein. 
Furthermore, there are attempts to generalize QFT to projective modules \cite{Gayral:2006wu} 
and to spectral geometries \cite{Paschke:2004xf}. Depending on the approach, one finds that 
NC quantum fields may exhibit new features, e.\,g.~a non-standard, so-called twisted, statistics.

In this work, we provide a generalization of the algebraic 
approach to QFT (see e.\,g.~\cite{Wald,baer} and references therein) to the realm of formally deformed
 manifolds, obtained by triangular Drinfel'd twists \cite{Drinfeld}. Even if twist-deformed 
(pseudo-)Riemannian manifolds may obey deformed isometry properties \cite{Ohl:2008tw}, we do not include
 these structures in the present work and focus on the QFT on in general nonsymmetric 
metric backgrounds. We will see later that the NC quantum field theory is described by an
(in general deformation parameter dependent) $\ast$-algebra of field observables, which is defined
by a symplectic structure. Thus, the advantage of our approach is that it remains close to 
 the standard algebraic setting of QFT. The obvious disadvantage 
 is the restriction to formal power series in the deformation parameter. This, in 
particular, does not allow us to study nonperturbative NC effects, such as causality violation in the
 field propagation. The main physical motivation for our investigations is to make contact to NC cosmology, 
in particular to perturbative NC effects in the cosmic microwave background, and to NC black hole physics. 
The NC gravity solutions recently obtained in \cite{Schupp:2009pt,Ohl:2009pv,Aschieri:2009qh} provide a 
natural application for the methods developed in this paper.

The outline of this paper is as follows: In Section \ref{sec:fields} we review the basics of twist-deformed 
differential geometry in the sense of \cite{Aschieri:2005yw,Aschieri:2005zs}. Based on these methods, we 
construct action functionals and equation of motion operators for free massive scalar fields for a 
large class of triangular Drinfel'd twists. In Section \ref{sec:solutions} we construct the deformed 
Green's operators and the deformed fundamental solution for the type of wave operators, constructed in
 Section \ref{sec:fields}, in terms of formal power series. The deformed wave propagation is causal with 
respect to the undeformed metric field. The deformed fundamental solution is used to canonically construct 
a symplectic vector space and to define $\ast$-algebras of field observables in Section \ref{sec:weyl}. 
We conclude and give an outlook to possible generalizations and applications of our formalism in Section \ref{sec:conc}.


\section{\label{sec:fields}Twist-deformed differential geometry, scalar field action functionals and equation of motion operators}

In this section we briefly describe how to construct action functionals and equation of motion 
operators for scalar fields on a large class of NC manifolds. We\index{\footnote{}} focus on the 
class of $\star$-products, which can be obtained by a Drinfel'd twist 
$\mathcal{F}\in U\Xi[[\lambda]] \otimes_\mathbb{C} U\Xi[[\lambda]]$, where $U\Xi$ is the universal
 enveloping algebra of the complexified vector fields on the manifold $\mathcal{M}$, $\lambda$ is the 
deformation parameter and $[[\lambda]]$ denotes formal power series in $\lambda$. The Drinfel'd twist 
should be finitely-generated, i.\,e.~$\mathcal{F}$ consists of sums of finite products of vector fields at 
every order in $\lambda$. For more information on these deformations see e.\,g.~\cite{Aschieri:2005zs}. 
Given the commutative algebra of smooth complex-valued functions $\mathcal{A}=\bigl(C^\infty(\mathcal{M}),\cdot\bigr)$ 
on the manifold $\mathcal{M}$, we can deform it into an associative, but in general noncommutative, 
algebra $\mathcal{A}_\star =\bigl(C^\infty(\mathcal{M})[[\lambda]],\star\bigr)$ with the $\star$-product
 defined by
\begin{flalign}
h \star k := \bar f^\alpha(h)\cdot \bar f_\alpha(k)~,~\text{for all }h,k\in\mathcal{A}_\star~,
\end{flalign}
where $\bar f^\alpha\otimes \bar f_\alpha = \mathcal{F}^{-1}$ is the inverse twist which acts on
 functions via the Lie derivative. We restrict ourselves in the following to hermitian $\star$-products 
satisfying $(h\star k)^\ast = k^\ast\star h^\ast$, for all $h,k\in\mathcal{A}_\star$, where 
$\ast$ is the standard involution on $\mathcal{A}$.

In the same way we can deform the exterior algebra of complexified differential forms 
$\bigl(\Omega^\bullet,\wedge,d\bigr)$ on $\mathcal{M}$ into $\bigl(\Omega^\bullet[[\lambda]],\wedge_\star,d\bigr)$ 
with the deformed wedge-product defined by
\begin{flalign}
\omega\wedge_\star \omega^\prime := \bar f^\alpha(\omega)\wedge \bar f_\alpha(\omega^\prime) ~,~\text{for all }\omega,\omega^\prime\in~\Omega^\bullet[[\lambda]]~,
\end{flalign}
where the inverse twist again acts on differential forms via the Lie derivative.
 Note that, since Lie derivatives and exterior derivatives commute, the exterior differential 
can be chosen to be undeformed and satisfies the graded Leibniz rule
\begin{flalign}
d\bigl(\omega\wedge_\star \omega^\prime\bigr) =( d\omega)\wedge_\star\omega^\prime + (-1)^{\text{deg}(\omega)} \omega\wedge_\star (d\omega^\prime)~,~\text{for all~}\omega,\omega^\prime\in\Omega^\bullet[[\lambda]]~.
\end{flalign}

Next, we consider the integration of differential forms on a twist-deformed manifold $\mathcal{M}$.
 Since, as vector spaces, the deformed and the formal power series of the undeformed differential forms are isomorphic,
 we can define integration in terms of the commutative integral. Identifying a volume
 form $\vol$ on $\mathcal{M}$ in general leads to a non-cyclic integral on $\mathcal{A}_\star$.
 Furthermore, it turns out that the integral for general $\mathcal{F}$ does not even possess the weaker 
property of ``graded cyclicity'' given by
\begin{flalign}
\int_\mathcal{M} \omega\wedge_\star \omega^\prime = (-1)^{\text{deg}(\omega) \text{deg}(\omega^\prime)}\int_\mathcal{M} \omega^\prime \wedge_\star \omega = \int_\mathcal{M} \omega\wedge \omega^\prime~,
\end{flalign}
for all $\omega,\omega^\prime\in \Omega^\bullet[[\lambda]]$ with 
$\text{deg}(\omega)+\text{deg}(\omega^\prime)=\text{dim}(\mathcal{M})$ and $\text{supp}(\omega)\cap\text{supp}(\omega^\prime)$
 compact \footnote{%
Let $\omega := \sum \lambda^n \omega_{(n)}\in\Omega^\bullet[[\lambda]]$ and $\omega^\prime := \sum \lambda^n \omega^\prime_{(n)}\in\Omega^\bullet[[\lambda]]$.
The statement $\text{supp}(\omega)\cap\text{supp}(\omega^\prime)$ compact is an abbreviation for
 $\text{supp}(\omega_{(n)})\cap\text{supp}(\omega^\prime_{(m)})$ compact for all $n,m\in\mathbb{N}^0$.
}. 
One explicit example of a twist not satisfying graded cyclicity is the Jordanian twist 
$\mathcal{F}=\exp\bigl(\frac{1}{2} H\otimes \log (1+\lambda E)\bigr)$ with $[H,E]=2 E$.
Since graded cyclicity is a property which drastically simplifies the 
construction of equations of motion from a given action functional, we restrict ourselves in the following 
to a subclass of Drinfel'd twists in order to obtain this property.
It can be checked that Drinfel'd twists satisfying $S(\bar f^\alpha)\cdot \bar f_\alpha = 1$
 fulfil graded cyclicity \cite{Aschieri:2009ky}. Here $S$ is the antipode on the universal enveloping algebra $U\Xi[[\lambda]]$, 
which is defined as an algebra antihomomorphism acting on vector fields $u\in\Xi[[\lambda]]$ by $S(u)=-u$ 
and $S(1)=1$. Note that the condition demanded above is not too restrictive, since it allows
 the whole class of abelian twists (also called Reshetikhin-Jambor-Sykora (RJS) twists \cite{Reshetikhin:1990ep,Jambor:2004kc}).

In order to construct kinetic terms of scalar field action functionals, we further require a deformed 
(pseudo-)hermitian structure $h_\star$ on one-forms. This structure can be defined using the inverse metric 
field of NC gravity $g^{-1}:= g^{-1\alpha}\otimes_\star g^{-1}_\alpha\in\Xi[[\lambda]]\otimes_\star\Xi[[\lambda]]$ 
and the pairing $\langle \cdot,\cdot\rangle_\star$ between vector fields $\Xi[[\lambda]]$ and one-forms 
$\Omega[[\lambda]]$ \cite{Aschieri:2005yw,Aschieri:2005zs} by
\begin{flalign}
h_\star(\omega,\omega^\prime) := \langle \langle \omega^\ast,g^{-1\alpha}\rangle_\star \star g^{-1}_\alpha,\omega^\prime\rangle_\star~,~\text{for all }\omega,\omega^\prime\in\Omega[[\lambda]]~.
\end{flalign}

Using the formalism described above, we are in the position to deform the standard quadratic scalar field 
action functional. We define
\begin{flalign}
\label{eqn:action}
S_\star[\Phi] := -\int_\mathcal{M} h_\star(d\Phi,d\Phi)\star \vol_\star - m^2 \int_\mathcal{M} \Phi^\ast\star\Phi\star\vol_\star~,
\end{flalign}
where $\vol_\star\in\Omega^\bullet[[\lambda]]$ is a real and nonvanishing top-form on $\mathcal{M}$ 
\footnote{
One consistent choice of $\vol_\star$ is the classical volume form associated to the NC metric field $g$, since it is an element of
 $\Omega^\bullet[[\lambda]]$, coordinate independent, real and nonvanishing on $\mathcal{M}$. Since we do not want to exclude different 
choices of $\vol_\star$ in our work, we keep it unspecified and only impose the natural conditions of reality and nondegeneracy.
}.
 Due to graded cyclicity, hermiticity of $h_\star$ and reality of $\vol_\star$ we obtain
$(S_\star[\Phi])^\ast = S_\star[\Phi]$.

The equation of motion for $\Phi$ is obtained by varying the action (\ref{eqn:action}). Since we 
are mainly interested in real scalar fields, we restrict ourselves from now on to this case.
Using variations $\delta\Phi$ of compact support and graded cyclicity we find
\begin{flalign}
\delta S_\star[\Phi]= \int_\mathcal{M} \delta \Phi \cdot \tilde P_\star[\Phi]~,
\end{flalign}
with the $\mathbb{C}[[\lambda]]$-linear top-form valued differential operator 
$\tilde P_\star:C^\infty(\mathcal{M})[[\lambda]]\to\Omega^\bullet[[\lambda]]$ given by
\begin{flalign}
\tilde P_\star[\varphi] := \tilde\square_\star[\varphi] + \bigl(\tilde\square_\star[\varphi^\ast] \bigr)^\ast - m^2 \varphi\star \vol_\star - m^2\vol_\star\star\varphi~,
\end{flalign}
for all $\varphi\in C^\infty(\mathcal{M})[[\lambda]]$.
The top-form valued d'Alembert operator $\tilde\square_\star$ is defined by
\begin{flalign}
\int_\mathcal{M} \psi^\ast \star \tilde\square_\star[\varphi] := - \int_\mathcal{M} h_\star(d\psi,d\varphi)\star \vol_\star~,
\end{flalign}
for all $\psi,\varphi\in C^\infty(\mathcal{M})[[\lambda]]$ with 
$\text{supp}(\psi)\cap\text{supp}(\varphi) $ compact.
Demanding $\delta S_\star[\Phi]=0$ results in the top-form valued equation of motion
 $\tilde P_\star[\Phi]=0$. By defining $\tilde P_\star[\varphi] =: P_\star[\varphi]\star \vol_\star$, 
for all $\varphi\in C^\infty(\mathcal{M})[[\lambda]]$, we obtain equivalently the equation of motion
 $P_\star[\Phi]=0$, where $P_\star:C^\infty(\mathcal{M})[[\lambda]]\to C^\infty(\mathcal{M})[[\lambda]]$ 
is now a $\mathbb{C}[[\lambda]]$-linear scalar differential operator.

Note that $P_\star$ is formally self-adjoint with respect to the scalar product
\begin{flalign}
\label{eqn:scalarproduct}
(\psi,\varphi)_\star := \int_\mathcal{M} \psi^\ast\star\varphi\star\vol_\star~.
\end{flalign}
More precisely, 
\begin{flalign}
\label{eqn:selfadjoint}
(\psi,P_\star[\varphi])_\star = (P_\star[\psi],\varphi)_\star~
\end{flalign}
holds true for all 
$\psi,\varphi\in C^\infty(\mathcal{M})[[\lambda]]$ with $\text{supp}(\psi)\cap\text{supp}(\varphi) $
 compact. Furthermore, in the case of a deformed Lorentzian manifold $(\mathcal{M},\star,g)$, 
the operator $P_\star$ is a formal deformation of a normally hyperbolic operator acting on 
$C^\infty(\mathcal{M})$, namely the Klein-Gordon operator.

This shows that we can explicitly construct quadratic scalar field action functionals leading 
to equation of motion operators which are formally self-adjoint with respect to the scalar product
 (\ref{eqn:scalarproduct}) and further are deformations of normally hyperbolic operators in the 
Lorentzian case. The generalization to generic matter fields requires the deformation quantization 
of hermitian vector bundles. For proofs on existence and uniqueness of these structures within the 
framework of formal deformation quantization see \cite{Waldmann}. The
explicit construction of realizations in 
twist-deformed NC geometry is currently in preparation \cite{AschieriSchenkel}.


\section{\label{sec:solutions}Formal solutions of deformed wave equations}
In this section we consider formal deformations of normally hyperbolic differential operators
 acting on formal power series of functions $C^\infty(\mathcal{M})[[\lambda]]$ on a twist-deformed
 globally hyperbolic  Lorentzian  manifold $(\mathcal{M},\star,g)$. This means that $(\mathcal{M},g\vert_{\lambda\to0})$
 is globally hyperbolic, but we do not impose conditions on the quantum corrections of $g$.
The physics which can be described within this framework are free scalar (quantum) field theories on a 
large class of curved NC manifolds. 

Let $P_\star :C^\infty(\mathcal{M})[[\lambda]]\to C^\infty(\mathcal{M})[[\lambda]]$ be a 
$\mathbb{C}[[\lambda]]$-linear differential operator, which is a formal deformation of a normally 
hyperbolic operator.  More precisely, $P_\star$ is defined by the formal power series
\begin{flalign}
 P_\star := \sum\limits_{n=0}^\infty \lambda^n P_{(n)}~,
\end{flalign}
where $P_{(0)}$ is a normally hyperbolic operator and 
$P_{(n)}:C^\infty(\mathcal{M})[[\lambda]]\to C_0^\infty(\mathcal{M})[[\lambda]]$ are the
 $\mathbb{C}[[\lambda]]$-linear ``quantum corrections'' for $n>0$.
 We demand that $P_\star$ is formally self-adjoint with respect to the scalar product (\ref{eqn:scalarproduct}), 
 i.\,e.~(\ref{eqn:selfadjoint}) holds true. We have constructed nontrivial examples of such operators in Section 
 \ref{sec:fields}, but we do not restrict ourselves to these examples in the following and work with the abstract definition.  
 
 Note that, in order to formally solve the dynamics governed by $P_\star$ we have to add one technical assumption.
 We assume that $P_{(n)}$ are differential operators which map to functions of compact support for $n>0$. 
 This is, similar to the interactions of compact support in perturbative QFT,
 a technical assumption required for the construction of the formal power series and not motivated by physics.   
 The support condition on $P_{(n)}$ is in particular satisfied for all twists constructed by compactly supported vector fields.
 In case of a noncompactly supported twist $\mathcal{F}$, an infrared (IR) regularization is in general required. 
 For this, we might, most preferably, try to approximate $\mathcal{F}$ by compactly supported twists or
 define by hand the regularized operators $P_{(n)}^\text{reg}:=\chi_{n}\, P_{(n)}$, where $\chi_n\in C_0^{\infty}(\mathcal{M})$
 are cutoff functions satisfying $\chi_n\vert_\mathcal{O}\equiv 1$ for a sufficiently large spacetime region $\mathcal{O}$.
 Although the global aspects of QFT are known to change due to IR regularization,
 it was argued in \cite{Duetsch:1998hf} that the local observables in (commutative) QFT remain unaffected
 by details of the IR regularization. Since our NC QFTs are formal deformations of commutative QFTs, we expect
 that, similar to \cite{Duetsch:1998hf}, the local formal NC physics is properly described by our models.

The main aim of this section is to construct deformed retarded and advanced Green's operators 
\begin{flalign}
 \Delta_{\star \pm}:= \sum\limits_{n=0}^\infty \lambda^n \Delta_{(n)\pm} : C_0^\infty(\mathcal{M})[[\lambda]]\to C^\infty(\mathcal{M})[[\lambda]]~
\end{flalign}
satisfying suitable properties. Even though some of the results obtained in this section could be derived 
from general considerations in deformation theory\footnote{We thank the anonymous referee for pointing this out to us.}, 
we perform an explicit construction of the deformed Green's operators
in order to make the investigation of their properties in this article self contained.

Theorems on the existence and uniqueness of Green's operators 
for the classical problem $\lambda\to0$ can be found in \cite{baer} and apply to the $\lambda^0$-order 
of our problem. In particular, it was shown in \cite{baer} that there exists an unique retarded 
and advanced Green's operator $\Delta_{\pm}=:\Delta_{(0)\pm} : C_0^\infty(\mathcal{M})\to C^\infty(\mathcal{M})$
 satisfying
\begin{subequations}
\label{eqn:greencl}
\begin{flalign}
 &P_{(0)} \circ \Delta_\pm = \text{id}_{C_0^\infty(\mathcal{M})}~,\\
 &\Delta_{\pm} \circ P_{(0)}\big\vert_{C_0^\infty(\mathcal{M})} = \text{id}_{C_0^\infty(\mathcal{M})} ~,\\
 &\text{supp}(\Delta_{\pm}[\varphi]) \subseteq J_\pm(\text{supp}(\varphi))~,\quad\text{for all~}\varphi\in C_0^\infty(\mathcal{M})~,
\end{flalign}
\end{subequations}
where $J_{\pm}(A)$ is the causal future/past of a subset $A$ measured with respect 
to the undeformed spacetime metric $g\vert_{\lambda\to0}$.

We extend these results to the NC setting by the following
\begin{theorem}
\label{theo:green}
 Let $(\mathcal{M},\star,g)$ be a deformed  time-oriented, connected, globally hyperbolic 
manifold and let $P_\star:=\sum\lambda^n P_{(n)}$ be a formal deformation of a normally 
hyperbolic operator acting on $C^\infty(\mathcal{M})[[\lambda]]$. Furthermore, let $P_{(n)}$ 
be finite-order differential operators with $\text{Im}(P_{(n)})\subseteq C_0^\infty(\mathcal{M})$ for $n>0$. 
Then there exist unique deformed Green's operators $\Delta_{\star\pm}:=\sum \lambda^n\Delta_{(n)\pm}$ satisfying
\begin{subequations}
\label{eqn:green}
\begin{flalign}
\label{eqn:green1}
& P_\star\circ\Delta_{\star\pm} = \text{id}_{C_0^\infty(\mathcal{M})[[\lambda]]}~,\\\label{eqn:green2}
& \Delta_{\star\pm} \circ P_\star\big\vert_{C_0^\infty(\mathcal{M})[[\lambda]]} = \text{id}_{C_0^\infty(\mathcal{M})[[\lambda]]}~,\\\label{eqn:green3}
& \text{supp}(\Delta_{(n)\pm}[\varphi])\subseteq J_{\pm}(\text{supp}(\varphi))~,\quad \text{for all~}n\in\mathbb{N}^0~\text{and~}\varphi\in C_0^\infty(\mathcal{M})~,
\end{flalign}
\end{subequations}
where $J_\pm$ is the causal future/past with respect to the classical metric $g\vert_{\lambda\to0}$.\\
The explicit expressions for $\Delta_{(n)\pm}$, $n>0$, read
\begin{multline}
\label{eqn:explicitgreen}
\Delta_{(n)\pm} = \sum\limits_{k=1}^{n}\sum\limits_{j_1=1}^n \dots\sum\limits_{j_k=1}^n(-1)^k \delta_{j_1+\dots+j_k, n} \\
  \Delta_{\pm}\circ P_{(j_1)}\circ \Delta_\pm \circ P_{(j_2)} \circ \dots \circ \Delta_\pm \circ P_{(j_k)}\circ \Delta_{\pm}~,
\end{multline}
where $\delta_{n,m}$ is the Kronecker-delta.
\end{theorem}
\noindent This theorem is proven in the Appendix \ref{app:green}.

Next, we study properties of the deformed fundamental solution defined by the $\mathbb{C}[[\lambda]]$-linear map
\begin{flalign}
\Delta_{\star} := \Delta_{\star +} - \Delta_{\star -}:C_0^\infty(\mathcal{M})[[\lambda]] \to C_\mathrm{sc}^\infty(\mathcal{M})[[\lambda]]~,
\end{flalign}
where $C_\mathrm{sc}^\infty(\mathcal{M})$ are the functions of spatially compact support. 
The importance of the fundamental solution lies in the fact that the covariant Poisson 
bracket relations (i.\,e.~the Peierls bracket relations) of classical field theory and
 the Weyl relations of QFT can be defined by using this map.
 We obtain the following
\begin{theorem}
\label{theo:complex}
 Let $(\mathcal{M},\star,g)$ be a deformed  time-oriented, connected, globally hyperbolic manifold 
and let $P_\star$ and $\Delta_{\star\pm}$ be as above. Then the sequence of $\mathbb{C}[[\lambda]]$-linear maps
\begin{flalign}
 0\longrightarrow C_0^\infty(\mathcal{M})[[\lambda]] \stackrel{P_\star}{\longrightarrow} C_0^\infty(\mathcal{M})[[\lambda]] \stackrel{\Delta_\star}{\longrightarrow} C_\mathrm{sc}^\infty(\mathcal{M})[[\lambda]] \stackrel{P_\star}{\longrightarrow}C_\mathrm{sc}^\infty(\mathcal{M})[[\lambda]]
\end{flalign}
is a complex, which is exact everywhere.
\end{theorem}
The proof of this theorem can be obtained by extending the proof of Theorem 3.4.7.~of 
\cite{baer} to formal power series. For completeness, we provide the proof in the Appendix \ref{app:complex}.


\section{\label{sec:weyl}Deformed symplectic vector space and $\ast$-algebras of field observables}
In the standard construction of the QFT of a free real scalar field on a 
commutative globally hyperbolic manifold with a normally hyperbolic operator $P$
one defines a symplectic structure $\omega$ 
on the real vector space $V:=C_0^\infty(\mathcal{M},\mathbb{R})/P[C_0^\infty(\mathcal{M},\mathbb{R})]$ 
by using the fundamental solution and the undeformed version of the scalar product 
(\ref{eqn:scalarproduct}). Using the symplectic vector space  $(V,\omega)$ one then defines 
the associated Weyl algebra of field observables. For details on this construction see \cite{baer}.

We now show that a similar construction is also possible in the twist-deformed case, 
leading to deformed $\ast$-algebras of field observables.
Consider a real scalar field 
on a deformed globally hyperbolic manifold with an equation of motion given by a deformed 
normally hyperbolic operator $P_\star$, satisfying the properties defined above. We further 
demand the reality condition of Section \ref{sec:fields} given by
\begin{flalign}
\label{eqn:realityeom}
 \bigl(\tilde P_\star[\varphi]\bigr)^\ast = \bigl(P_\star[\varphi]\star \vol_\star \bigr)^\ast = \tilde P_\star[\varphi^\ast]~,~~\text{for all~}\varphi\in C^\infty(\mathcal{M})[[\lambda]]~,
\end{flalign}
since it naturally emerges from an action principle.

Using Theorem \ref{theo:green} we obtain unique deformed Green's operators $\Delta_{\star\pm}$ 
and the fundamental solution $\Delta_\star= \Delta_{\star+}-\Delta_{\star-}$. We define the following map
\begin{flalign}
\label{eqn:preomega}
\tilde\omega_\star: C^\infty_0(\mathcal{M})[[\lambda]] \times C^\infty_0(\mathcal{M})[[\lambda]] \to\mathbb{C}[[\lambda]],~(\psi,\varphi)\mapsto\tilde\omega_\star(\psi,\varphi)= (\psi,\Delta_\star[\varphi])_\star~,
\end{flalign}
which is the basic ingredient in the construction of the deformed symplectic vector space $V_\star$.
In order to define $V_\star$, we first have to restrict $C^\infty_0(\mathcal{M})[[\lambda]]$ 
to a suitable real subspace. We define the real vector space \begin{flalign}
 H :=\lbrace \varphi\in C^\infty_0(\mathcal{M})[[\lambda]]: (\Delta_{\star\pm}[\varphi])^\ast =\Delta_{\star\pm}[\varphi] \rbrace~.
\end{flalign}
This vector space turns out to be a natural generalization of the classical space
 $C_0^\infty(\mathcal{M},\mathbb{R})$ due to the following
\begin{proposition}
\label{propo:Hspace}
Consider the vector space $H$ defined above. Then the following statements hold true:
\begin{itemize}
\item[1.)] Let $\psi\in C_\mathrm{sc}^\infty(\mathcal{M},\mathbb{R})[[\lambda]]$ be a real solution of the wave equations given by $P_\star$. Then there is a $\varphi\in H$, such that $\psi = \Delta_\star[\varphi]$.
\item[2.)] The kernel of the fundamental solution $\Delta_\star$ restricted to $H$ is given by $\mathrm{Ker}(\Delta_\star)\vert_H = P_\star[C^\infty_0(\mathcal{M},\mathbb{R})[[\lambda]]]$.
\item[3.)] Let $\varphi\in H$, then $(\varphi \star \vol_\star)^\ast = \varphi\star\vol_\star$.
\end{itemize}
\end{proposition}
\noindent This proposition is proven in the Appendix \ref{app:Hspace}.

We define the real factor space $V_\star:=H/P_\star[C_0^\infty(\mathcal{M},\mathbb{R})[[\lambda]]]$. 
The map $\tilde\omega_\star$ (\ref{eqn:preomega}) induces a map $\omega_\star$ on $V_\star$ by defining
\begin{flalign}
\omega_\star:V_\star\otimes_\mathbb{R} V_\star \to\mathbb{R}[[\lambda]],~([\psi],[\varphi])\mapsto \tilde\omega_\star(\psi,\varphi)=(\psi,\Delta_\star[\varphi])_\star~.
\end{flalign}
This map is well defined due to the anti-hermiticity property 
$(\psi,\Delta_\star[\varphi])_\star = -(\Delta_\star[\psi],\varphi)_\star$ for 
all $\psi,\varphi\in C^\infty_0(\mathcal{M})[[\lambda]]$, which follows from Lemma 
\ref{lem:antihermiticity} (see Appendix \ref{app:complex}).
Furthermore, $\omega_\star$ is nondegenerate due to Proposition \ref{propo:Hspace}, part $2.)$,
and the nondegeneracy of the scalar product.

It remains to show the antisymmetry and reality of this map.
Using Proposition \ref{propo:Hspace}, we obtain for all $\psi,\varphi\in H$
\begin{multline}
(\psi,\Delta_\star[\varphi])_\star = \int_\mathcal{M} \psi^\ast \star \Delta_\star[\varphi]\star \vol_\star \stackrel{\text{GC},\text{RE}}{=} \int_\mathcal{M} (\psi\star\vol_\star)^\ast \star \Delta_\star[\varphi] \stackrel{3.)}{=} \\
\int_\mathcal{M} \psi\star\vol_\star \star \Delta_\star[\varphi] \stackrel{\text{GC}}{=} \int_\mathcal{M} \Delta_\star[\varphi] \star\psi\star\vol_\star \stackrel{\varphi\in H}{=}(\Delta_\star[\varphi],\psi)_\star \stackrel{\text{AH}}{=} - (\varphi,\Delta_\star[\psi])_\star~,
\end{multline}
where we also have used graded cyclicity (GC), reality of $\vol_\star$ (RE) and 
anti-hermiticity of $\Delta_\star$ (AH). From this identity we obtain that $\omega_\star$ 
is antisymmetric. Reality follows from
\begin{flalign}
(\psi,\Delta_\star[\varphi])_\star^\ast \stackrel{\text{HSP}}{=} (\Delta_\star[\varphi],\psi)_\star \stackrel{\text{AH}}{=} -(\varphi,\Delta_\star[\psi])_\star \stackrel{\text{AS}}{=} (\psi,\Delta_\star[\varphi])~,
\end{flalign}
where we have used hermiticity of the scalar product (HSP), anti-hermiticity of 
$\Delta_\star$ and antisymmetry of $\omega_\star$ (AS).

This shows that, given a twist-deformed real scalar field with equation of motion 
operator $P_\star$ defined as above, we can construct the symplectic vector space 
$(V_\star,\omega_\star)$. Using this vector space we can, guided by the commutative case \cite{baer},
define an $\ast$-algebra of field observables as follows:
\begin{definition}
\label{defi:Weyl}
 Let $(V_\star,\omega_\star)$ be a symplectic vector space. Let $\mathfrak{A}$ be an $\ast$-algebra
 over $\mathbb{C}[[\lambda]]$ with unit and let $W:V_\star\to\mathfrak{A}$ be a map, such that for all 
 $\varphi,\psi\in V_\star$ we have
 \begin{subequations}
 \begin{flalign}
  &~~W(0)=1~,\\
  &~~W(-\varphi)=W(\varphi)^\ast~,\\
  &~~W(\varphi)\cdot W(\psi) = e^{-i \omega_\star(\varphi,\psi)/2}~ W(\varphi+\psi)~.
 \end{flalign}
\end{subequations}
 We call $\mathfrak{A}$ an $\ast$-algebra of Weyl-type, if it is generated by 
 the elements $W(\varphi)$.
\end{definition}
Note that, different to the commutative case, we did not demand $\mathfrak{A}$ to be a $C^\ast$-algebra,
since we are considering algebras over $\mathbb{C}[[\lambda]]$ and not $\mathbb{C}$. 
See \cite{Waldmann2} for a review on $\ast$-algebras over ordered rings and their $\ast$-representation 
theory on pre-Hilbert spaces. It is well-known that the uniqueness (up to $\ast$-isomorphisms) of 
the Weyl algebra in commutative QFT strongly relies on the $C^\ast$-property. Thus, we expect a richer 
$\ast$-representation theory for the $\ast$-algebras of Weyl-type defined above. However, in case we would 
find a convergent deformation of the symplectic vector space $(V_\star,\omega_\star)$, what is strongly
 motivated by physics, we could define the Weyl system according to the conventional definition
 \cite{baer}, including $C^\ast$-algebras.

A second possible definition of an $\ast$-algebra of field observables is the following:
\begin{definition}
\label{defi:Poly}
 Let $(V_\star,\omega_\star)$ be a symplectic vector space. Let $\mathfrak{A}$ be an $\ast$-algebra
over $\mathbb{C}[[\lambda]]$ with unit and let $\Phi:V_\star\to\mathfrak{A}$ be a $\mathbb{R}[[\lambda]]$-linear map, 
such that for all $\varphi,\psi\in V_\star$ we have
\begin{subequations}
\begin{flalign}
&~~\Phi(\varphi)^\ast=\Phi(\varphi)~,\\
&~~[\Phi(\varphi),\Phi(\psi)]= i\,\omega_\star(\varphi,\psi)\, 1~.
\end{flalign}
\end{subequations}
We call $\mathfrak{A}$ an $\ast$-algebra of field polynomials, if it is generated 
by the elements $1$ and $\Phi(\varphi)$.
\end{definition}
This definition of the $\ast$-algebra of field polynomials is closer to physics, since there the 
$n$-point correlation functions, i.e.~expectation values of $n$-fold products of the linear field operators
in some appropriate state, are of particular interest. This shows that there are natural definitions of
field observable algebras in the NC setting, motivating the algebraic description of QFT on curved NC spacetimes.
The explicit construction and investigation of these algebras is beyond the scope of our present work.

We conclude this section by briefly studying the observable algebras of a very special class of 
twist-deformations of the manifold $\mathcal{M}$. Let $(\mathcal{M},g,\vol)$ be a Lorentzian manifold with
 isometries given by a Lie algebra $\mathfrak{g}$ and with the
 volume form~$\vol$ associated to~$g$. Furthermore, let $\mathcal{F}\in U\mathfrak{g}[[\lambda]]\otimes_\mathbb{C}U\mathfrak{g}[[\lambda]]\subseteq U\Xi[[\lambda]]\otimes_\mathbb{C}U\Xi[[\lambda]]$.
 These Drinfel'd twists are called Killing twists, since they are constructed completely
 by Killing vector fields. It is easy to see that  all $\star$-products drop out of the action (\ref{eqn:action})
by using graded cyclicity and the $\mathfrak{g}$-invariance of $\vol$ and~$g$.
Thus, the equation of motion operator $P_\star = P$ is undeformed, as well as the fundamental
 solution $\Delta_\star=\Delta$. We further obtain that $V_\star =V$ and that the symplectic structure 
$\omega_\star=\omega$ is undeformed, leading to the undeformed observable algebras. This means that the 
free QFT on Killing deformed manifolds can not be distinguished from the commutative
 one. Of course, interacting QFTs on Killing deformed 
spacetimes will contain NC effects. Note that the Moyal-Weyl deformation of the Minkowski 
spacetime is a particular example of a Killing deformation, since the twist-generating 
vector fields $\lbrace P_\mu := \partial_\mu\rbrace$ are Killing.


\section{\label{sec:conc}Conclusions and outlook}
In this paper we have made a first step towards generalizing the algebraic approach to QFT 
to the realm of formally deformed manifolds. The deformations we have considered are given by triangular 
Drinfel'd twists, together with the restriction $S(\bar f^\alpha)\cdot\bar f_\alpha =1$, and include 
in particular all abelian (also called RJS) twists. We have constructed quadratic scalar field actions
 and equations of motion within the framework of twist-deformed differential geometry and have shown that 
these wave equations can be solved in terms of formal power series. More precisely, we have constructed 
deformed advanced and retarded Green's operators and the deformed fundamental solution explicitly up 
to all orders in the deformation parameter. The deformed wave propagation is, as expected, compatible 
with classical causality given by the metric field $g\vert_{\lambda\to0}$, since we have considered formal
 deformations. Using the fundamental solution, we were able to construct a symplectic vector space and
therewith define $\ast$-algebras of field observables for the NC quantum field, 
which in general depend on the deformation parameter $\lambda$. In the special case of the Moyal-Weyl 
deformed Minkowski spacetime, or more general for all Killing deformations of symmetric manifolds, 
the deformation parameter is absent in the observable algebra.

In future work it will be of particular importance to study the physics described by the 
$\ast$-algebras of field observables using explicit examples, e.\,g.~cosmological models or black holes. 
For this purpose the NC gravity solutions recently obtained in \cite{Schupp:2009pt,Ohl:2009pv,Aschieri:2009qh} 
will be helpful. It would also be fruitful to study the differences to other approaches to NC QFT 
\cite{NCQFT} within these models.

On the conceptual side, it would be interesting to include the ideas of locally covariant QFT 
\cite{Brunetti:2001dx} to the NC setting, see also \cite{Paschke:2004xf} for such an 
attempt. If this is possible at all, and what is the role of the twisted diffeomorphisms 
\cite{Aschieri:2005yw,Aschieri:2005zs} within this approach, are issues left to future work.


\begin{acknowledgements}
We want to thank Paolo Aschieri and Christoph Uhlemann for discussions and comments on this work.
This research is supported by Deutsche Forschungsgemeinschaft through the Research 
Training Group GRK\,1147 \textit{Theoretical Astrophysics and Particle Physics}.
\end{acknowledgements}

\appendix
\section{\label{app:green}Proof of Theorem \ref{theo:green}}
 We perform a proof by induction. The zeroth order of (\ref{eqn:green}) is assured
 by (\ref{eqn:greencl}). Assume that we have constructed the Green's operators 
to order $\lambda^{n-1}$. In order $\lambda^n$ (\ref{eqn:green1}) reads
\begin{flalign}
\label{eqn:pr1}
 \sum_{m=0}^n P_{(m)}\circ \Delta_{(n-m)\pm}=0~.
\end{flalign}
Let $\varphi\in C_0^\infty(\mathcal{M})$ be arbitrary. We can reformulate 
(\ref{eqn:pr1}) into a Cauchy problem with respect to the
 classical operator $P_{(0)}$
\begin{flalign}
 P_{(0)}[\Delta_{(n)\pm}[\varphi]] = - \sum\limits_{m=1}^n P_{(m)}[\Delta_{(n-m)\pm}[\varphi]]=:S\in C_0^\infty(\mathcal{M})~.
\end{flalign}
In order to satisfy (\ref{eqn:green3}), we have to impose trivial Cauchy data of vanishing field and derivative 
on a Cauchy surface past/future to $\text{supp}(S)$. 
The unique solution is
\begin{flalign}
\label{eqn:sol}
 \Delta_{(n)\pm}[\varphi] =\Delta_{\pm}[S]= -\sum\limits_{m=1}^n \Delta_{\pm}\circ P_{(m)}\circ \Delta_{(n-m)\pm}[\varphi]~.
\end{flalign}
We obtain the support property for $n \geqq m>0$
\begin{multline}
 \text{supp}(\Delta_{\pm}\circ P_{(m)}\circ \Delta_{(n-m)\pm}[\varphi] )\subseteq J_{\pm}(\text{supp}(P_{(m)}\circ\Delta_{(n-m)\pm}[\varphi]))\\
\subseteq J_{\pm}(\text{supp}(\Delta_{(n-m)\pm}[\varphi]))\subseteq  J_{\pm}(J_{\pm}(\text{supp}(\varphi))) \subseteq J_{\pm}(\text{supp}(\varphi))~,
\end{multline}
where we have used that $P_{(m)}$ is a finite-order $\mathbb{C}$-linear differential operator, 
thus satisfying $\text{supp}(P_{(m)}[\varphi])\subseteq \text{supp}(\varphi)$.
 This shows that (\ref{eqn:green3}) is satisfied to order $\lambda^n$.

The equality of (\ref{eqn:sol}) and (\ref{eqn:explicitgreen}) can either be 
shown combinatorically, or by showing that (\ref{eqn:explicitgreen}) solves 
(\ref{eqn:green1}) together with the support property (\ref{eqn:green3}), and thus has to 
be equal to (\ref{eqn:sol}).

The remaining step is to prove the order $\lambda^n$ of (\ref{eqn:green2}). 
Plugging in the explicit form (\ref{eqn:explicitgreen}) one notices that every 
possible chain of operators, e.\,g.~
\begin{flalign}
 \Delta_{\pm}\circ P_{(j_1)} \circ \Delta_{\pm}\circ \dots \circ \Delta_{\pm}\circ P_{(j_k)}~,\quad \text{with~} j_1+j_2+\dots+j_k=n~,
\end{flalign}
occurs exactly twice in (\ref{eqn:green2}), but with a different sign. Thus they cancel.\qed

\section{\label{app:complex} Proof of Theorem \ref{theo:complex}}
In order to prove Theorem \ref{theo:complex} we require the following two lemmas.
\begin{lemma}
\label{lem:scsupport}
 Let $\varphi\in C^\infty_0(\mathcal{M})[[\lambda]]$, then $\Delta_{\star}[\varphi]\in C^\infty_\mathrm{sc}(\mathcal{M})[[\lambda]]$.
\end{lemma}
\begin{proof}
 Define $\psi = \sum \lambda^n \psi_{(n)} := \Delta_{\star}[\varphi]$. We obtain by using Theorem \ref{theo:green}
\begin{flalign}
 \text{supp}(\psi_{(n)}) \subseteq J_+(K^\varphi_{(n)})\cup J_-(K^\varphi_{(n)})~,~~\text{for all~}n\in\mathbb{N}^0~,
\end{flalign}
where  $K^\varphi_{(n)} := \bigcup_{m=0}^n\text{supp}(\varphi_{(n-m)})$ is compact. 
Thus $\psi_{(n)}\in C^\infty_\mathrm{sc}(\mathcal{M})$ for all $n\in\mathbb{N}^0$.\qed
\end{proof}
\begin{lemma}
 \label{lem:antihermiticity}
$(\psi,\Delta_{\star\pm}[\varphi])_\star = (\Delta_{\star\mp}[\psi],\varphi)_\star$ for all $\psi,\varphi\in C^\infty_0(\mathcal{M})[[\lambda]]$.
\end{lemma}
\begin{proof}
 By Theorem \ref{theo:green} we have
\begin{flalign}
 (\psi,\Delta_{\star\pm}[\varphi])_\star = (P_\star\circ \Delta_{\star\mp}[\psi],\Delta_{\star\pm}[\varphi])_\star~.
\end{flalign}
Using Theorem \ref{theo:green} and global hyperbolicity, one obtains that $\text{supp}(\Delta_{\star\mp}[\psi])\cap\text{supp}(\Delta_{\star\pm}[\varphi])$ 
is compact for all $\psi,\varphi\in C^\infty_0(\mathcal{M})[[\lambda]]$. To finish the proof 
we use that $P_\star$ is formally self-adjoint.\qed
\end{proof}

We now give a proof of Theorem \ref{theo:complex}.
\begin{proof}
The sequence of maps forms a complex due to Theorem \ref{theo:green} and Lemma \ref{lem:scsupport}. 

To prove the first exactness, let $\varphi\in C^\infty_0(\mathcal{M})[[\lambda]]$ such that $P_\star[\varphi]=0$. 
Then $\varphi = \Delta_{\star\pm}\circ P_\star[\varphi] = 0$. 

To prove the second exactness, let $\varphi\in C^\infty_0(\mathcal{M})[[\lambda]]$ such that $\Delta_{\star}[\varphi]=0$. 
We define $\psi:=\Delta_{\star\pm}[\varphi]$ and obtain using Theorem \ref{theo:green} 
and global hyperbolicity of $(\mathcal{M},g\vert_{\lambda\to0})$ 
that $\psi\in C^\infty_0(\mathcal{M})[[\lambda]]$. We find $P_\star[\psi]=P_\star\circ\Delta_{\star\pm}[\varphi]=\varphi$.

To prove the third exactness, let $\varphi=\sum\lambda^n\varphi_{(n)}\in C^\infty_\mathrm{sc}(\mathcal{M})[[\lambda]]$ such that $P_\star[\varphi]=0$.  
We can find a family of compact sets $\lbrace K_{(n)}: n\in\mathbb{N}^0\rbrace$, such that
 $\text{supp}(\varphi_{(n)})\subseteq I_+(K_{(n)})\cup I_-(K_{(n)})$, where $I_{\pm}(A)$ is
 the chronological future/past of a subset $A$ with respect to the classical metric $g\vert_{\lambda\to0}$.
 We decompose analogously to \cite{baer} $\varphi_{(n)}=\varphi_{(n)}^+ +\varphi_{(n)}^-$, where 
$\text{supp}(\varphi_{(n)}^\pm)\subseteq I_{\pm}(K_{(n)})\subseteq J_\pm(K_{(n)})$. We define 
$\varphi^{\pm} := \sum\lambda^n\varphi_{(n)}^\pm$ and $\psi:= \pm P_\star[\varphi^\pm]$. 
Using the support properties of $\varphi^\pm$ and global hyperbolicity, one obtains that $\psi\in C^\infty_0(\mathcal{M})[[\lambda]]$.
To show that $\Delta_{\star\pm}[\psi]=\pm\varphi^{\pm}$, let $\chi\in C^\infty_0(\mathcal{M})[[\lambda]]$ be arbitrary. We obtain
\begin{flalign}
 (\chi,\Delta_{\star\pm}[\psi])_\star = (\Delta_{\star\mp}[\chi],\psi)_\star = \pm (\Delta_{\star\mp}[\chi],P_\star[\varphi^\pm])_\star = \pm(P_\star\circ\Delta_{\star\mp}[\chi],\varphi^\pm)_\star = (\chi,\pm \varphi^\pm)_\star~,
\end{flalign}
where we have used Lemma \ref{lem:antihermiticity}, Theorem \ref{theo:green} and that $P_\star$ is formally self-adjoint. 
This shows that $\Delta_{\star}[\psi] = \varphi$.\qed
\end{proof}

\section{\label{app:Hspace}Proof of Proposition \ref{propo:Hspace}}
{\it Proof of 1.):}\\
Let $\psi\in C^\infty_\mathrm{sc}(\mathcal{M},\mathbb{R})[[\lambda]]$ be a real solution 
satisfying $P_\star[\psi]=0$. By Theorem \ref{theo:complex} we know that there is a
 $\varphi\in C^\infty_0(\mathcal{M})[[\lambda]]$, such that $\psi = \Delta_\star[\varphi]$.
 From the reality of $\psi$ we obtain
\begin{flalign}
 \bigl(\Delta_{\star+}[\varphi]\bigr)^\ast - \Delta_{\star+}[\varphi] = \bigl(\Delta_{\star-}[\varphi]\bigr)^\ast - \Delta_{\star-}[\varphi]=:2 \delta\in C_0^\infty(\mathcal{M})[[\lambda]]~.
\end{flalign}
One obtains that $\delta$ is of compact support by using Theorem \ref{theo:green} and global hyperbolicity.
Using $\delta^\ast = -\delta$ and $\delta = \Delta_{\star\pm}\circ P_\star[\delta]$ 
(see Theorem \ref{theo:green}) we find that 
\begin{flalign}
 \left(\Delta_{\star\pm}\left[\varphi + P_\star[\delta]\right]\right)^\ast = \Delta_{\star\pm}\left[\varphi + P_\star[\delta]\right]~.
\end{flalign}
Thus $\varphi+P_\star[\delta]\in H$ with $\Delta_\star[\varphi+P_\star[\delta]] = \Delta_{\star}[\varphi]=\psi$. \qed
~\\
\noindent {\it Proof of 2.):}\\
Let $\varphi\in P_\star[C^\infty_0(\mathcal{M},\mathbb{R})[[\lambda]]]$, then there is a 
$\chi\in C^\infty_0(\mathcal{M},\mathbb{R})[[\lambda]]$, such that $\varphi=P_\star[\chi]$. We obtain 
\begin{flalign}
 (\Delta_{\star\pm}[\varphi])^\ast = \chi^\ast =\chi =\Delta_{\star\pm}[\varphi]~.
\end{flalign}
Thus $\varphi\in H$ and $\Delta_\star[\varphi] = \Delta_{\star}[P_\star[\chi]]=0$.\\
Let now $\varphi\in H$ such that $\Delta_\star[\varphi]=0$. Then by Theorem 
\ref{theo:complex} there exists $\chi\in C^\infty_0(\mathcal{M})[[\lambda]]$, such that
 $\varphi=P_\star[\chi]$. Using the definition of $H$ we obtain
\begin{flalign}
 0=\left(\Delta_{\star\pm}[\varphi]\right)^\ast-\Delta_{\star\pm}[\varphi] = \chi^\ast -\chi~,
\end{flalign}
thus $\chi\in C^\infty_0(\mathcal{M},\mathbb{R})[[\lambda]]$. \qed
~\\
\noindent {\it Proof of 3.):}\\
Let $\varphi\in H$. Using reality of the top-form valued equation of motion operator
 (\ref{eqn:realityeom}) we obtain
\begin{flalign}
 \left(\varphi\star\vol_\star\right)^\ast = \left(\tilde P_\star\left[\Delta_{\star\pm}\left[\varphi\right]\right]\right)^\ast = \tilde P_\star\left[\left(\Delta_{\star\pm}\left[\varphi\right]\right)^\ast \right]= \tilde P_\star\left[\Delta_{\star\pm}\left[\varphi\right] \right]=\varphi\star\vol_\star~.
\end{flalign}
\qed


\end{document}